%Paper: hep-th/9204070
%From: SPALLUCCI%TRIESTE.INFN.IT@icnucevm.cnuce.cnr.it
%Date: Wed, 22 APR 92 13:52 N
%Date (revised): Fri, 24 APR 92 09:53 N

%%%%%%%%%%%%%%%%%%%%%%%%%%%%%%%%%%%%%%%%%%%%%%%%%%%%%%%%%%%%%%%%%
\input phyzzx
\mathsurround=2pt
\openup 2pt
\hfuzz 30pt
\pubnum{UTS-DFT-92-11}
\pubtype{HE}
\titlepage
\singlespace
\def\a{anomaly}
\def\as{anomalies}
\def\a{anomaly}
\def\as{anomalies}

\def\sc{Schwinger}

\title{Gravitational and Schwinger model anomalies: how far can
the analogy go?}

\author{A.Smailagic\foot{E-mail address: ANAIS@ITSICTP.BITNET}}
\address{International Center for Theoretical Physics, Trieste, Italy\break
Strada Costiera 11-34014, Trieste, Italy}
\andauthor{E.Spallucci\foot{E-mail address: SPALLUCCI@TRIESTE.INFN.IT}}
\address{ Dipartimento di Fisica Teorica\break Universit\`a di
Trieste\break
I.N.F.N., Sezione di Trieste\break
Trieste, Italy}
\submit{ Phys.Lett.B}
\doublespace
\endpage
%\vfill\eject
\abstract

We describe the most general treatment of all anomalies
both for chiral and massless Dirac fermions, in two-dimensional
gravity.
It is shown that for this purpose two regularization dependent
parameters are present in the effective action.
Analogy to the \sc\ model is displayed corresponding to a specific
choice of the second parameter, thus showing that the gravitational
model contains \a\ relations having no analogy in the \sc\ model.

\vfill\eject

\REFS\Aa{%
A.Smailagic, Phys.Lett.{\bf B195}, (1987), 213
}%
\REFSCON\Ab{%
A.Smailagic, Phys.Lett.{\bf B205}, (1988), 483
}%
\REFSCON\Ac{%
A.Smailagic, R.E.Gamboa-Saravi, Phys.Lett.{\bf B192}, (1987), 145
}%
\REFSCON\Ad{%
P.Breitenlohner, D.Maison, Commun.Math.Phys.{\bf 52}, (1977), 39,55\hfill\break
G.Thompson, H.L.Yu, Phys.Lett.{\bf B151}, (1985), 119;\hfill\break
H.L.Yu, W.B.Yeung Phys.Rev.{\bf D39}, (1987), 636
}%
\REFSCON\Ae{%
L.Alvarez-Gaume, E.Witten, Nucl.Phys.{\bf B234}, (1983), 269
}%
\REFSCON\Afuji{%
K.Fujikawa, Phys.Rev.Lett.{\bf 42}, 1195 (1979); Phys.Rev.{\bf D21}, 2848
(1980);
}%
\REFSCON\Ag{%
H.Leutwyler,Phys.Lett.{\bf B153}, (1985), 65
}%
\REFSCON\Aproc{%
A.Smailagic, E.Spallucci ``~Chiral Models in Gravity~'', in Proceedings
of the ``~XVII International Colloquium on Group Theoretical Methods
in Physics~'', ed. World Sci. 1988
}%
\REFSCON\Ah{%
L.N.Chang, H.T.Nieh, Phys.Rev.Lett.{\bf 53}, (1984), 21
}%
\REFSCON\Ajj{%
R.Jackiw, R.Rajaraman, Phys.Rev.Lett.{\bf 54}, (1985), 1219
}%
\REFSCON\Ak{%
C.M.Hull, P.K.Townsend, Nucl.Phys.{\bf B274}, (1986), 349
}%
\REFSCON\Anoi{%
A.Smailagic, J.A.Helayel-Neto, Mod.Phys.Lett.{\bf A10}, (1987),
787;\hfill\break
A.Smailagic, E.Spallucci ``~Lorentz Anomaly in $(1,0)$ Induced Supergravity~''
UTS-DFT-92-6, submitted to Phys.Lett.{\bf B}
}%

\refsend

Two dimensional models of massless fermions coupled to gravity have
been extensively
used as toy-models for the understanding of gauge and gravitational \as.
A variety of method of calculations has been developed and various specific
models have been studied. Apparently, two distinct situations are present:
models with either Dirac or Weyl fermions. The former presents the freedom
of preserving one of the classical symmetries at the expense of the other,
while in the latter anomalies are {\it always} present.
This apparent discrepancy has been explained, and a unified description
of \as\  in both cases has been given in the case of the Schwinger
model\refmark{\Aa}. A preliminary discussion according to such a vantage
viewpoint, in the case
of gravity, has been given as well\refmark{\Ab}. However, the resulting
effective action contained a multitude of arbitrary parameters since
the local parts have been constructed in all possible ways in terms of
zweibein, and, though
giving correct results, an analogy to the Schwinger model was not evident.
Furthermore, one has to bear in mind that in the case of fermions
coupled to gravity one has
{\it three} classical symmetries to start with, while in the case of the
Schwinger model one encounters only two gauge symmetries. Thus, it is
reasonable to expect, but necessary to verify
explicitly, that it will not be possible to parametrize all the regularization
ambiguities in terms of a single parameter, as it results from ref.(2).
 Notwithstanding, the classical analogy between the \sc\ model and fermions
in a two-dimensional curved spacetime is compelling. The
results of ref.(1) show that the zweibein is not a proper choice as a gauge
field bearing similarity to the gauge potential $A_\mu$. Another possible
choice is the spin-connection $\omega^{ab}{}_\mu$ that in two dimensions
can be always written as
$$
\omega^{ab}{}_\mu=\epsilon^{ab}\omega{}_\mu\ .
\eqno(1)
$$
On the other hand, the spin-connection is expressed in terms of the zweibein
through the vanishing torsion constraint. In two-dimensions its
{\it linearized} form, $e^a{}_\mu=\delta^a{}_\mu +h^a{}_\mu$, is given
%%%%%%%%%%%%%%%%%%%%%%%%%%%%%%%%%%%%%%%%%%%%%%%%%%
by\foot{ We maintain at the linearized
level too the mixed notation with latin and greek indices,
even if it is not strictly necessary because in this approximation
all the indices are flat. This is to avoid
confusion with the ``~standard~'' notation, where the graviton field
$h_{\mu\nu}=g_{\mu\nu}-\eta_{\mu\nu}$ is {\it symmetric} by definition,
while in our case the linearized zweibein $h_{a\mu}$ has no definite
symmetry property under $a\leftrightarrow\mu$ exchange.}

%%%%%%%%%%%%%%%%%%%%%%%%%%%%%%%%%%%%%%%%%%%%%%%%%%%%%%%%
$$
\omega_\mu=\epsilon^{ab}\partial_b\bar h_{a\mu}+{1\over2}\epsilon_{\mu a}
\partial^a h +{1\over2}\partial_\mu L
\eqno(2)
$$
where we decompose the ``~graviton~'' field in traceless symmetric part
$\bar h_{a\mu}$, totally anti-symmetric part $h_{[a\mu]}$, and trace
$h$, according to
$$
h_{a\mu}=\bar h_{a\mu}+{1\over2}\delta_{a\mu}h+h_{[a\mu]}\ ,
\eqno(3)
$$
finally, $L=\epsilon^{ab}h_{ab}$ and $h=h_{a\mu}\delta^{a\mu}$ are the Lorentz
and Weyl degrees of freedom respectively.

{}From eq.(2) we learn that:

\noindent
i) in two dimensions $\omega_\mu$
should have only two components, therefore the decomposition in (2) shows
that some of the components must be gauge degrees of freedom. Actually, at
the classical level all of them can be gauged away since gravity has no
dynamics. On the quantum level this indicates that, if anomalies are present,
{\it they combine together in such a way that $\omega_\mu$ has
at most two dynamical components.}

\noindent
ii) Comparing (2) to the decomposition of the gauge field $A_\mu$ for the
Schwinger model
$$
A_\mu=\partial_\mu \phi+\epsilon_{\mu\nu}\partial^\nu\rho
\eqno(4)
$$
we expect a close analogy between gravitational and Schwinger model only in
the case
$\bar h_{a\mu}=0$, which amounts to imposing general coordinate invariance
at the quantum level.

With these general arguments in mind, we are going to perform an explicit
anomaly calculation starting from the action
$$
W_{\rm F}={i\over 2}\int d^2x\,{\rm e} e_a{}^\mu\bar\psi\gamma^a
\left(\partial_\mu -\omega_{\mu\,ab}\Sigma^{ab}\right)
\left({1-\beta\gamma^5\over2}\right)\psi+{\rm H.C.}\ ,
\eqno(5)
$$
where ${\rm e}\equiv{\rm det}e^a{}_\mu$, and the parameter $\beta$ has been
introduced to be able to consider simultaneously both Dirac($\beta=0$)
and Weyl fermions($\beta=\pm 1$).

To clarify our arguments we shall first compute the effective action in
a manifestly general covariant scheme and introduce the regularization
parameter not ``~by hand~'', i.e. as the coefficient of a local counter-term,
but rather through the formal properties of the two-dimensional Dirac
operator\refmark{\Ac}.
Then, we shall extend the above scheme to the case where all the three
symmetries are broken at the quantum level, and introduce a second arbitrary
parameter.

Various method of calculation are at our disposal. One would naturally tend
to perform perturbative calculations. However, if one is to keep $\omega_\mu$
as a genuine gauge field throughout the calculations one is faced with
the problem of vanishing spin current in two dimensions, leading
to the lack of a ``~current-gauge field~'' vertex. Then, either
one continues $\gamma^5$ out of two dimensions\refmark{\Ad}, which has
created some confusion in the literature on how useful these prescription
are for an explicit calculation, or works indirectly in terms
of ``~zweibein-energy momentum tensor~'' vertices\refmark{\Ae}
and reconstructs the spin-connection in the final results.
To make the long story short and transparent we opt for the Fujikawa
approach\refmark{\Afuji}. The basis of this algorithm is a proper choice of the
regularizing operator of the Jacobian arising from the
symmetry transformations of the functional measure.
As an appropriate choice we choose as a regularizing operator in Minkowski
space\foot{
We shall adopt the following conventions for the (minkowskian) Dirac matrices:
$$
\{\gamma^a,\gamma^b\}=2\eta^{ab}\ ,\eta_{00}=+1\ ,\eta_{11}=-1\ ;\quad
\gamma^5={1\over2}\epsilon^{ab}\gamma_a\gamma_b\ .
$$
}
$$
D=ie_m{}^\mu\gamma^m\partial_\mu +e_m{}^\mu\gamma^5\gamma^m\omega_\mu\left(
{1-\beta\gamma^5\over2}\right)\ .
\eqno(6)
$$
Following the usual procedure,  we obtain, from eq.(6), the euclidean
form of the Fujikawa regulator as
$$
D_E=ie_m{}^\mu\gamma^m\partial_\mu +{1\over2}e_m{}^\mu\gamma^m\Omega^5{}_\mu+
{i\over2}e_m{}^\mu\gamma^5\gamma^m\Omega_\mu\ ,
\eqno(7)
$$
where
$$\eqalign{
\Omega^5{}_\mu&\equiv
\left(ag_{\mu\nu}+{\beta\over2}\sqrt{-g}\epsilon_{\mu\nu}
\right)\omega^{5\,\nu}\ ,\cr
\Omega_\mu&\equiv\left((1-a)g_{\mu\nu}+{\beta\over2}\sqrt{-g}\epsilon_{\mu\nu}
\right)\omega^\nu\ ,\cr}
\eqno(8)
$$
and we have introduced an arbitrary parameter $a$ exploiting
the two-dimensional identity
$\gamma^\mu\gamma^5= \epsilon^{\mu\nu}\gamma_\nu$.
The final result of such a procedure, after analytic continuation
to imaginary time of the Dirac operator, is a splitting of $D_E$ into the
sum of an hermitian and a non-hermitian part. Although such a splitting
may seem straightforward it may be done in different ways (at least in
the case of chiral fermions) and one has to be careful in order to find
the correct anomalies. The introduction of an arbitrary parameter $a$
represents, in the Fujikawa approach, the counterpart of perturbative
calculation ambiguities corresponding to different choices of the
regularization method.

Using known formulae\refmark{\Afuji} we can obtain the anomalous Ward
identities
for Weyl and Lorentz symmetries as
$$\eqalignno{
T&=i{\rm tr}\left(A_1\left[D_E^\dagger D_E\right]-
A_1\left[D_ED_E^\dagger\right]\right)\ ,
&(9a)\cr
T_{[\mu\nu]}&=-{i\over2}\sqrt{-g}\epsilon_{\mu\nu}
{\rm tr}\left(\gamma^5 A_1\left[D_E^\dagger D_E\right]+
\gamma^5 A_1\left[D_ED_E^\dagger\right]\right)\ , &(9b)\cr}
$$
where $A_1\left[\Delta\right]$ is the second coefficient of the
heat-kernel expansion
for the elliptic, second order, differential operator $\Delta$.
The above formulae show the need to
consider non-hermitian operators in the Fujikawa approach in order to obtain
all the anomalies. The choice of  the hermitian part of the operator
in eq.(7) corresponds to the Weyl
invariant regularization, while the choice of an anti-hermitian Dirac operator
corresponds to the Lorentz invariant regularization.  An arbitrarily
weighted combination of the two gives both anomalies.

Rotating back to Minkowski spacetime we obtain\foot{
For simplicity we have rescaled all the following formulae by
the global numerical coefficient appearing in the effective action as
$1/192\pi$.
}
$$
\eqalignno{
T&={1\over\sqrt{-g}}\epsilon^{\mu\nu} \partial_\mu \Omega_\nu &(10a)\cr
T_{[\mu\nu]}&=-\epsilon_{\mu\nu}\epsilon^{\rho\sigma}\partial_\rho
\Omega^5{}_\sigma
\ .&(10b)\cr}
$$
With the help of eq.(8), eq.(10) can be written in a more familiar and
\sc-looking form
$$\eqalignno{
T&=2(1-a)R+\beta \partial_\mu \omega^\mu &(11a)\cr
T_{[\mu\nu]}&=-{1\over2}\epsilon_{\mu\nu}\left(\beta R+2a\partial_\mu\omega^\mu
\right)\ . &(11b)\cr}
$$
{}From the explicit expression of the anomalies one can construct the
corresponding
effective action. The easiest way to do it, is to notice that the trace and the
anti-symmetric part of the energy-momentum tensor couple to the trace
and to the anti-symmetric part of the zweibein and then expressing the later
two in terms of spin-connection. In this way, one finds the effective
action with the help of eqs.(2) and (11):
$$
I^{\rm eff}=
\int d^2x \sqrt{-g}\left[R\,{1\over\nabla^2}
\left(R+\beta\partial_\mu\omega^\mu
\right)+ag^{\mu\nu}\omega_\mu\omega_\nu \right]\ ,
\eqno(12)
$$
where $\nabla^2=\nabla_\mu\nabla^\mu$ is the scalar covariant D'Alambertian.
Before we proceed let us make a few comments on the obtained results.

First of all, as previously explained, we have considered a {\it particular}
situation in which the anomalies of {\it only} Lorentz and Weyl symmetry
are considered, while general coordinate invariance is assumed to be valid
at the quantum level. Clearly, this is {\it not} the most general situation,
but certainly the one that exhibits analogy to the \sc\ model.

Secondly, comparison to \sc\ model effective action shows an asymmetric
form of the non-local part in eq.(12). It is our believe that this has
created a certain confusion in the literature\refmark{\Ag}
and we would like to explain
this apparent difference. It is possible to show that there is a
relation between the symmetric and the asymmetric form of the effective
actions. This is due to the relation
$$
R^2=\left(\nabla_\mu\omega^\mu\right)^2-
\nabla_\mu\nabla^\mu\omega_\nu\omega^\nu
\eqno(13)
$$
which leads to
$$\eqalign{
&R\,{1\over\nabla^2}\left(R+{2\beta\over 1+\beta^2}\nabla_\mu\omega^\mu\right)+
b\,\omega_\mu\omega^\mu\equiv\cr
&\left({1\over 1+\beta^2}\right)\left[\left(R+\beta\nabla_\mu\omega^\mu\right)
{1\over\nabla^2}\left(R+\beta\nabla_\mu\omega^\mu\right)+a\omega_\mu\omega^\mu
\right]\cr}
\eqno(14)
$$
where the two arbitrary parameters $a$ and $b$ are related by $\displaystyle{b=
{a+\beta^2\over 1+\beta^2}}$. Both (12) and (15) lead to the same anomalies
 once the above relation between the parameters is taken into account.
Then,
both form of the effective action can be used on the same footing, and
(12) can be rewritten, in complete analogy to the
\sc\ model, as
$$
I^{\rm eff}=\int d^2x \sqrt{-g}\,\omega_\mu
\left[ag^{\mu\nu}-\left({1\over\sqrt{-g}}
\epsilon^{\mu\rho}+\beta g^{\mu\rho}\right)\nabla_\rho\,{1\over\nabla^2}
\,\nabla_\sigma\left({1\over\sqrt{-g}}
\epsilon^{\sigma\nu}+\beta g^{\sigma\nu}\right)\right]\omega_\nu\ .
\eqno(15)
$$
As we hope to have explained at the beginning, and confirmed through our
explicit calculation, the analogy to the \sc\ model (i.e. arbitrariness in
terms of a single parameter) is possible as long as we consider {\it only}
Lorentz and Weyl symmetries, {\it assuming} quantum general covariance.
We would like to go further and consider all three symmetries on the same
footing, i.e. without making any assumptions on their validity at the
quantum level.
To do so we need to devise a way of breaking general covariance.
Since all the components of the zweibein transform under general
coordinate transformations, contrary to Weyl and Lorentz symmetry, the presence
of all the components in covariant expressions is a necessary condition
to guarantee
covariance, and the absence of one of them will spoil it. Hence, it
seems the most convenient to introduce  an {\it additional} parameter $c$
in the decomposition (3) as
$$
h^{(c)}{}_{a\mu}=
\bar h_{a\mu}+{c\over2}\left(\delta_{a\mu}h+2h_{[a\mu]}\right)\ .
\eqno(16)
$$
We shall not proceed by any explicit choice of regularization but rather
consider the problem in its full generality.

One finds linearized transformations of the zweibein (16) as
$$
\delta h^{(c)}{}_{a\mu}={c+1\over2}\partial_\mu\xi_a+
{1-c\over2}\partial_a\xi_\mu+{c-1\over2}\delta_{a\mu}\partial_d\xi^d+
c\Lambda\delta_{a\mu}-c\theta\epsilon_{a\mu}\ .
\eqno(17)
$$
where $\xi^\mu$,$\Lambda$ and $\theta$ are covariant, Weyl and Lorentz
transformation parameters.
Further, using eq.(2) in its linearized form, we obtain the following
variations
$$\eqalignno{
&\delta\omega^{(c)}{}_\mu={1-c\over2}\epsilon^{de}\partial_e\partial_\mu\xi_d+
{c-1\over2}\epsilon_{\mu d}\partial^d\partial_a\xi^a+
c\epsilon_{\mu a}\partial^a\Lambda-c\partial_\mu \theta &(18a)\cr
&\delta\partial^\mu\omega^{(c)}{}_\mu=
{1-c\over2}\epsilon^{de}\partial^2\partial_e\xi_d-c\partial^2 \theta &(18b)\cr
&\delta\epsilon^{\nu\mu}\partial_\nu\omega^{(c)}{}_\mu=
{c-1\over2}\partial^2\partial_d\xi^d+c\partial^2\Lambda\equiv-\delta
R^{(c)}\ . &(18c)\cr}
$$
{}From eq.(18c) one can
see that the linearized variation of the Ricci scalar with respect to
a general coordinate transformation is zero for the choice $c=1$ (which leads
to the decomposition (3)).
{\it If $c\ne 1$ general covariance is spoiled}
as expected from the above discussion. With the generalized
choice (16), starting from the symmetric form of the effective action
(15), and using eq.(18) one can find anomaly relations for general covariance,
Lorentz and Weyl symmetries as\foot{Although we used linearized variations (18)
at the end of calculations we have reconstructed covariant results.
One could have used the Fujikawa approach with the decomposition (17), which
gives the same anomalies (19).}
$$
\eqalignno{
\nabla^\nu T_{\mu\nu}&=(1-c)\Biggl\{\nabla_\mu\left[(1-a)R+\beta\nabla_\rho
\omega^\rho\right]+\cr
&\>\>\>\sqrt{-g}\epsilon_{\mu\rho}\nabla^\rho\left[(a+\beta^2)
\nabla_\sigma\omega^\sigma+\beta R\right]\Biggr\} &(19a)\cr
T&=-c\left[(1-a)R+\beta \nabla_\mu \omega^\mu\right] &(19b)\cr
T_{[\mu\nu]}&={c\over 2}\epsilon_{\mu\nu}\left[\beta R+(a+\beta^2)
\nabla_\rho\omega^\rho\right]\ . &(19c)\cr}
$$
It is evident
from eqs.(19) that introduction of the second parameter was crucial
to put all the \as\ of the quantum gravity on the same footing
and, therefore, to describe the most general situation.
It contains
all the partial results present in the literature as specific choices
of the constants $a,\,c,\,\beta$, and gives further
insight in the interplay among various anomalies in quantum gravity.

First of all, one notices the emergence of the general pattern conjectured
at the beginning, and
based on the spin-connection decomposition, i.e. either one chooses to
preserve general covariance ($c=1$) introducing
Weyl and Lorentz anomalies, or breaks general covariance preserving
Weyl and Lorentz symmetry\refmark{\Aproc} ($c=0$).
In the former case one can see that
eqs.(19) leads to the anomaly relations (11) and the analogy to the \sc\ model
is due to the absence of the covariant anomaly.

Within this general scheme there is still a further freedom related to the
choice of the parameters $a$ and $\beta$. It is generally believed that
in models with Dirac fermions the interplay of \as\ is between general
covariance and Weyl symmetry, preserving Lorentz symmetry.

This view needs some explanation based on the results in eqs.(19):
as already said,
parameter $c$ is relevant for absence of general covariant \a\ while
parameter $a$ interpolates (once $c$ is chosen) between Lorentz and Weyl \as.
Therefore, insisting on the absence of general covariant \a\ one can, in
principle, still have Lorentz \a\ and no Weyl \a\ ($a=0$).
However, the opposite
is usually preferred ($a=1$) {\it but not necessary}, and surely, absence
of covariant \a\ {\it does not} imply a priori absence of Lorentz \a.

The case of chiral fermions ($\beta=\pm 1$) is even more intriguing since
no choice of the parameter $a$ can remove Lorentz
or Weyl \as\ (similarly to the \sc\ model), which are, therefore,
genuine quantum effects\refmark{\Ab,\Ah}, and not regularization dependent
pathologies.

Various possibilities exist in this case. One can always
produce combinations of $T\pm \epsilon^{ab}T_{ab}$ which are proportional to
either $1+\beta$ or $1-\beta$, and choosing, say $\beta=1$, only one of them
survives. The analogy with the chiral \sc\ model, where the anomalous
current is the combination of the gauge and axial currents and no choice
of the parameter can eliminate the \a\ of this combined current\refmark{\Aa},
is then evident. This remark is
relevant as long as one wants to write the induced effective action in terms
of only one light-cone component of the gauge field, i.e. the one that is
classically coupled to chiral fermions. Otherwise, one can keep the local
term ($a\ne 0$) in which both light-cone components of the gauge field
are present\refmark{\Ajj}.

In the case of the spin-connection, the situation is a bit more
general than in the \sc\ model, and we feel it deserves a detailed explanation.
Chiral models are best described in terms of light-cone components and from
now on we shall stick to this notation\foot{ The scalar product is written
in terms of light-cone components as
$$
A_\mu B^\mu\equiv {1\over2}(A_+B_-+A_-B_+)\ .
$$
}.
The decomposition (2) can be written as
$$\eqalignno{
\omega_-&=\partial_+\bar h_{--}+{1\over 2}\partial_-(L-h) &(20a)\cr
\omega_+&=\partial_-\bar h_{++}+{1\over 2}\partial_+(L+h) &(20b)\cr}
$$
and eqs.(20) can be compared to the light-cone decomposition of the gauge
field $A_\mu$ for the \sc\ model
$$
A_\pm=\partial_\pm(\phi\pm\rho)\ .
\eqno(21)
$$
The scheme we previously described for chiral models corresponds to having
only $\omega_+$ or $\omega_-$ ($a=0$). From eq.(20) it is possible to
have only one component of the spin-connection, in the chiral models, by
using residual symmetries to impose, say $\bar h_{\pm\pm}=0$ and $L=-h$,
thus obtaining the anomaly equation
$$
T_{-+}=\partial_+\omega_-\ .
\eqno(22)
$$
Keeping in mind that $T_{-+}\equiv(\delta_{ab}-\epsilon_{ab})\partial^a S^b=
\partial_-S_+$, with $S^a$ the quantum spin-current, the preceding
relation can be written as
$$
\partial_-S_+=\partial_+\omega_-\ ,
\eqno(23)
$$
which is apparently the \sc\ model anomaly with substitution $S_+\rightarrow
J_+$ and $\omega_-\rightarrow A_-$\refmark{\Aa}.

The other possibility corresponds to choose $L=0$ and $h=0$ independently
(by preserving both Lorentz
and Weyl symmetry) and $\bar h_{++}=0$, (choice of particular
chiral coupling $a=0\ ,\quad\beta=1$) producing a gravitational \a\refmark{\Ae}
$$
\partial_-T_{++}=\partial^2_+\omega_-\equiv\partial^3_+\bar h_{--}\ .
\eqno(24)
$$
{\it
This is an  equation that has no analogue in the case of the \sc\ model.}
Both situations can be simply described in terms of an anomalous equation
with $\omega_-$ given by eq.(20a) and, in one case $\bar h_{--}$ is the
dynamical variable, while in the other case it is $L-h$.

We have shown that the general treatment of all \as\ of two-dimensional
quantum gravity
requires {\it two} arbitrary (regularization dependent) parameters and,
therefore, analogy to \sc\ model is achieved only with the particular
choice of the second parameter. Within a particular regularization
scheme this parameters are fixed and, for example, $c=0$ corresponds to
the two-dimensional decomposition of the gravitational coupling in the
Feynman rules ${\rm e}e_{a\mu}=\delta_{a\mu}+\bar h_{a\mu}$, leading to
the covariant \a\refmark{\Ae}, while $c=1$ is achieved by
${\rm e}e_{a\mu}=\delta_{a\mu}+{2-D\over 2}\delta_{a\mu}h
+\bar h_{a\mu} +h_{[a\mu]}$\refmark{\Ak}.
The fixing of the parameter $a$ is already known
through the \sc\ model literature. Spin-connection turns out to be more
appropriate as a gauge field, not only because it bears similarity
to the vector gauge potential of the \sc\ model, but also because it
leads to the quantum spin current and nicely separate local from
non-local pieces in the effective action, allowing ,in such a way,
the identification of
arbitrary parameters (and reducing their number). It is still worth
reminding that we are working within the second order formulation and
treat spin-connection as a function of zweibein.

Within present formulation gravitational (in terms of $\omega_\mu$), vector
an axial (in terms of $A_\mu$) \as\ of the \sc\ model are treated
separately in terms of independent regularization parameters. It would be
interesting to see what kind of relation among parameters is induced
in super-symmetric models of gravity, particularly an $(2,2)$ super-gravity
where axial gauge field is a member of a gravitational super-multiplet.
Preliminary investigation in this direction, but with different motivations,
has been done both for $(0,1)$ and $(1,1)$
super-gravity\refmark{\Anoi} without giving answer to the above question
since the axial gauge field, in these models, is not yet a member of the same
super-multiplet as the graviton and gravitino. This point is now under
investigation.

\refout
\bye